\documentclass[sigconf]{acmart} 
\usepackage{balance}

\def\BibTeX{{\rm B\kern-.05em{\sc i\kern-.025em b}\kern-.08emT\kern-.1667em\lower.7ex\hbox{E}\kern-.125emX}}

\usepackage[utf8]{inputenc}
\usepackage{hyperref}
\usepackage{url}
\usepackage[]{algorithm2e}
\usepackage{amssymb}
\usepackage{graphicx}
\usepackage{booktabs}
\usepackage{url}
\usepackage{tabu}
\usepackage{xcolor}

\usepackage{hyperref}
\usepackage{underscore}
\usepackage{tcolorbox}
\newcommand*{\Scale}[2][4]{\scalebox{#1}{$#2$}}%

\newcolumntype{P}[1]{>{\centering\arraybackslash}p{#1}}

\usepackage{standalone}

\copyrightyear{2019} 
\acmYear{2019} 
\setcopyright{acmcopyright}
\acmConference[SIGIR '19]{Proceedings of the 42nd International ACM SIGIR Conference on Research and Development in Information Retrieval}{July 21--25, 2019}{Paris, France}
\acmBooktitle{Proceedings of the 42nd International ACM SIGIR Conference on Research and Development in Information Retrieval (SIGIR '19), July 21--25, 2019, Paris, France}
\acmPrice{15.00}
\acmDOI{10.1145/3331184.3331307}
\acmISBN{978-1-4503-6172-9/19/07}

\author{Casper Hansen}
\affiliation{
  \city{University of Copenhagen}
}
\email{c.hansen@di.ku.dk}
\author{Christian Hansen}
\affiliation{
  \city{University of Copenhagen}
}
\email{chrh@di.ku.dk}
\author{Stephen Alstrup}
\affiliation{
  \city{University of Copenhagen}
}
\email{s.alstrup@di.ku.dk}
\author{Jakob Grue Simonsen}
\affiliation{
  \city{University of Copenhagen}
}
\email{simonsen@di.ku.dk}
\author{Christina Lioma}
\affiliation{
  \city{University of Copenhagen}
}
\email{c.lioma@di.ku.dk}

\title{Contextually Propagated Term Weights for Document Representation }

\begin{document}

\begin{abstract}
Word embeddings predict a word from its neighbours by learning small, dense embedding vectors. In practice, this prediction  corresponds to a semantic score given to the predicted word (or term weight). We present a novel model that, given a target word, redistributes part of that word's weight (that has been computed with word embeddings) across words occurring in similar contexts as the target word. Thus, our model aims to simulate how semantic meaning is shared by words occurring in similar contexts, which is incorporated into bag-of-words document representations. Experimental evaluation in an unsupervised setting against 8 state of the art baselines shows that our model yields the best micro and macro F1 scores across datasets of increasing difficulty.
\end{abstract}

\maketitle

\section{Introduction}
\label{s:intro}
\textit{Word embeddings} represent words as elements in a learned vector space by mapping semantically similar words to nearby points (otherwise put, by \textit{embedding} them nearby each other). By doing so, word embeddings adopt the \textit{Distributional Hypothesis}, which posits that words appearing in the same contexts share semantic meaning \cite{LiomaSLH15}. 
An efficient and popular implementation for learning word embeddings is word2vec \cite{mikw2w} and the skip-gram model. Given a target word, skip-gram is trained to predict the words in the context of that target word. This prediction practically corresponds to a score (or \textit{weight}) given to the predicted word, which can be applied to a range of tasks.

We present a novel model\footnote{The code is available at: \url{https://github.com/casperhansen/CPTW}} that, given a target word, redistributes part of that word's weight (that has been computed with word embeddings) back to words occurring in similar contexts as the target word. By doing so, 
our model aims to simulate how semantic meaning is shared by words occurring in the same context, by sharing (or \textit{propagating}) the semantic scores computed for these words within the neighbourhood of contextually similar words. This propagation is incorporated into bag-of-words document representations. 
We experimentally evaluate our model in the task of unsupervised text clustering against 3 established and 5 state of the art baselines. Our model yields the best micro and macro $\textrm{F}_1$ scores on average across all datasets. In addition, our model is efficient and robust across datasets of different inter versus intra class ratio (i.e. across datasets of increasing difficulty).

\section{Related Work}
\label{s:rw}
A simple, yet often effective practice of outputting document scores from word embeddings is by the average of the embedding of each word in a text \cite{weavg}. A disadvantage of this approach is that all words are weighted equally, failing to distinguish between more and less discriminative words. 
Arora et al.\ \cite{sif} proposed an unsupervised word embedding weighting scheme using word occurrence probabilities from a large corpus. Each embedded word is weighted by:
$\alpha/(\alpha + p(w))$ with $p(w)$ being the probability of word $w$ occurring, and $\alpha$ being a smoothing parameter. From each averaged vector the first singular vector is subtracted, which corresponds to removing common words (e.g. \texttt{just}, \texttt{when}, and \texttt{even}) \cite{sif}. The method of Arora et al.\ was defined and evaluated for sentences.
Kusner et al.\ \cite{wmd} recently presented the unsupervised word mover's distance (WMD), which approximates the semantic distance between two documents by computing the earth mover's distance between the embedded words in each document. 
While WMD experimentally outperformed several state of the art methods in the task of text clustering, its main disadvantage is that it is computationally very costly and cannot be readily used on medium to large scale datasets.


\section{Contextually Propagated Term Weights}
\label{s:cptw}
\label{ourapproach}
Given a word embedding \cite{mikolov2013efficient}, we define as the \textit{embedded neighbourhood} of term $w_j$ in document $d_i$, the set of all similar terms having cosine similarity to $w_j$ at least $\tau$ in the embedding space (i.e. the most similar terms thresholded by a minimum cosine similarity of $\tau$), where $\tau$ is a tunable threshold value (discussed in Section \ref{sec:threp})\footnote{Note that \textit{embedded neighbourhood} is not the same as \textit{embedding space}: the former is derived from the similarities measured in the latter.}. We use this embedded neighbourhood to define two contextual document representations, presented next.

\subsection{CPTW}
Let $N(w_j)$ be the set of words contained in the embedded neighbourhood of word $w_j$ (note that this includes $w_j$ itself). 
Then, for each $w_k \in N(w_j)$, we define:
\begin{equation}
\label{eq:sptw}
\gamma(w_k) = f(w_k, d_i) \cos(v_j, v_k)
\end{equation}
\noindent where $\gamma(w_k)$ is our contextually propagated term weight of $w_k$, $f(w_k,d_i)$ is the frequency of word $w_k$ in document $d_i$, and $\cos(v_j,v_k)$ is the cosine similarity between the word embeddings for word $w_j$ and $w_k$. Eq. \ref{eq:sptw} computes the term weight of $w_j$ and of each word in $w_j$'s embedded neighbourhood. 
Then, based on the above term weights, we compute the contextually propagated term weights (CPTW) of document $d_i$, denoted $\textrm{CPTW}(d_i)$, as:
\begin{equation}
\label{eq:cptw}
\textrm{CPTW}(d_i) = \sum_{j=1}^M e_j \Big( \alpha_j \sum_{w_k \in N(w_j)} \gamma(w_k) \Big)
\end{equation}
\noindent where 
$M$ is the total number of unique words in the collection (such that the representation resides in $\mathbb{R}^M$), 
$e_j$ is the vector with $1$ at index $j$ and zero everywhere else, 
$\gamma(w_k)$ is the term weight of $w_k$ computed using Eq. \ref{eq:sptw}, and
$\alpha_j$ is a normalization constant, computed as:
$
\alpha_j = \left( \sum_{w_k \in N(w_j)} \cos(v_j, v_k) \right)^{-1}
$. In Eq. \ref{eq:cptw}, $\alpha_j$ ensures that all words have the same total weight independent of the number of words they are similar to. This has the benefit that a word with a larger embedded neighbourhood (larger $N(w_j)$) is not weighted higher than a word with a smaller embedded neighbourhood (smaller $N(w_j)$). When the number of unique words is considered fixed then a $\tau$-tresholded word-to-word similarity matrix can be computed offline, such that the similarity propagation of CPTW can be trivially done efficiently using sparse vector-matrix multiplications on a traditional bag-of-words representation. Fig. \ref{fig:example} shows an example of CPTW in practice.

\subsection{$\textrm{CPTW}_\textrm{IDF}$}
Eq. \ref{eq:sptw} approximates the weight of term $w_k$ according to (i) the frequency of its occurrence in $d_i$ ($f(w_k, d_i)$), and (ii) how similar $w_k$ is to  $w_j$ ($\cos(v_j, v_k)$). The frequency of $w_k$ in $d_i$ can be artificially inflated for terms that occur very often in the collection, not just in $d_i$ (just like TF in traditional bag of words computations when not combined with IDF). To counter this effect, we introduce the following variation of $\gamma(w_k)$ that includes an IDF-like component, thus ensuring that high within-document term frequency \textbf{and} term discriminativeness in the collection are considered when computing term weights:

\begin{equation}
\label{eq:sptw-idf}
\gamma_{IDF}(w_k) = \gamma(w_k) \log \left( \frac{N}{\text{df}(w_k)} \alpha_j \cos(v_j, v_k) \right) 
\end{equation}
\noindent where 
$\gamma(w_k)$ is computed as in Eq. \ref{eq:sptw}, 
$\text{df}(w_k)$ is the number of documents in the collection that contain word $w_k$, 
$N$ is the total number of documents in the collection, and 
the rest of the notation is the same as above.
Note that in this case $\alpha_j$ is placed inside the $\log$ because we propagate the IDF component and still need the normalization for the same reason as above.
We define the $\textrm{CPTW}_\textrm{IDF}$ of document $d_i$ based on the above term weights as:
\begin{equation}
\label{eq:cptw-idf}
\text{CPTW}_{\text{IDF}}(d_i) = \sum_{j=1}^M e_j \Big( \alpha_j  \sum_{w_k \in N(j)} \gamma_{IDF}(w_k) \Big)
\end{equation}
\noindent The purpose of the IDF component is the same as in traditional TF-IDF; however, in our approach it is computed by propagating the IDF values for each word in the embedded neighbourhood as done similarly for term frequency in Eq. \ref{eq:cptw}. This means that the IDF score for each word is based on a weighted sum of all the IDF scores of the words in its embedded neighbourhood, thus taking into account the differences in IDF scores between the words in the embedded neighbourhood.






\subsection{Threshold parameter $\tau$\label{sec:threp}}
The optimal value of $\tau$ depends on both (i) the quality of the word embedding (to avoid dissimilar words from being represented as similar), and (ii) the similarity of the texts in the collection.
If the collection consists of texts with vastly different topics, then a low $\tau$ is preferred because there is little overlap of word semantics between texts with different topics. However, if the topics are similar, then $\tau$ should be higher to avoid including semi-related words that are shared across topics.
If, for the sake of explanation, we assume that no thresholding is done, then each word contains all words in its embedded neighbourhood, and thus the value of each word in the bag-of-words vector would be the sum of cosine similarities to all other words. As the cosine similarity is a dot product of unit length vectors, the distributive property of the dot product entails that the sum of cosine similarities would be, in effect, the dot product of the embedded word vector and the sum of \emph{all} other embedded word vectors. In order to not consider this sum of all other embedded word vectors, we choose $\tau$ to define a smaller neighbourhood such that the sum only consists of words that are deemed similar enough to not introduce noise (i.e. minuscule or negative dot products).

\begin{figure}
\begin{tcolorbox}
\hspace{-15pt}
\begin{minipage}{.59\linewidth}
\scalebox{0.89}{1: The {\color{blue}{boat}} is {\color{blue}{sailing}} on the {\color{blue}{sea}}} \\
\scalebox{0.89}{2: The {\color{blue}{ship}} was {\color{blue}{cruising}} on the {\color{blue}{ocean}}}\\
\scalebox{0.89}{3: The cat was relaxing on the couch}
\end{minipage}
\begin{minipage}{.39\linewidth}
\scalebox{0.89}{
\begin{tabular}{l|c|c|c}
& \multicolumn{3}{c}{Euclidean distance} \\ 
     & 1$\leftrightarrow $2 & 1$\leftrightarrow $3 & 2$\leftrightarrow $3 \\ \hline
 BOW &  0.40 &  0.40  & 0.35  \\ \hline
 CPTW & 0.16 & 0.49 & 0.40
\end{tabular}
}
\vspace{-9pt}
\end{minipage}
\end{tcolorbox}
\vspace{-10pt}
\caption{\label{fig:example}Example: Our similarity propagation reduces the Euclidean distance between semantically similar texts compared to bag-of-words with frequency weighting (BOW).}
\vspace{-15pt}
\end{figure}

\section{Experimental Evaluation}
\label{s:eval}
We experimentally evaluate our contextually propagated term weights (CPTW) and CPTW${_\text{IDF}}$ against strong baselines for related document representation and document distance methods that use few to zero parameters, applied to text clustering. 
For all methods we apply k nearest neighbour (kNN) classification for evaluation purposes, in order to purely focus on the representation that each method can generate to discriminative between texts.

\subsection{Data}
We use 7 openly available datasets commonly used in related work \cite{wmd,kenter2015short,gupta2016doc2sent2vec,zhang2016bayesian}. As seen in  Table \ref{table:data}, the datasets are largely varied with respect to their vocabulary size, number of unique words in each document, and number of classes.


\begin{table}[h]
    \centering
    \scalebox{0.8}{
\begin{tabular}{|l|c|c|c|c|}
\hline
Dataset & \#docs & unique words & \#avg. unique words $\pm$ std. & \#classes \\ \hline 
bbcsport  & 737 & 13106 & 116.4$\pm$55.9 & 5 \\ 
twitter   & 3424  & 8405 &  8.9$\pm$3.3 &  3 \\ 
classic  & 7095  & 23624  & 39.3$\pm$28.1 &  4\\ 
amazon  & 8000  & 39852  & 44.47$\pm$46.9  & 4\\ 
reuters  & 7674  & 23109  & 38.9$\pm$36.9  & 8\\ 
20news & 18846 & 30465 & 84.1$\pm$96.9 & 20 \\
wiki & 19981 & 46610 & 474.9$\pm$411.7 & 25 \\ 
\hline
\end{tabular}}
\caption{\label{table:data}Data statistics.}
\vspace{-30pt}
\end{table}

\subsection{\label{refmethods}Baselines and Tuning}
We compare our CPTW and CPTW$_{\text{IDF}}$ models against bag-of-words with frequency weighting (BOW) \cite{salton1988term}, TF-IDF \cite{salton1988term}, BM25 \cite{robertson1995bm25}, LSI \cite{deerwester1990lsi}, LDA \cite{blei2003lda}, Word Embedding Averaging (WE-AVG) \cite{weavg}, Smooth inverse frequency weighting (SIF) \cite{sif}, and Word Mover's Distance (WMD) \cite{wmd}.
To tune and evaluate all models we use 5-fold cross validation, where each fold acts as testing data once, while the rest is used for training and validation. In each iteration, we use 70\% of the 4 folds for training and the remaining 30\% for validation. We tune the parameters of all models using a grid search of the parameter space and repeat this 3 times in each iteration (the best parameters across the 3 iterations are chosen for the test fold). For \textit{reuters}, \textit{20news}, and \textit{wiki} we use the existing splits.

We use the micro and macro versions of $\textrm{F}_1$ for evaluation, and for validation we use the micro $\textrm{F}_1$ (corresponding to traditional accuracy) for choosing the best parameters. In all methods we search kNN's $k\in\{1,...,19\}$ as done in a similar setup by Kusner et al.\ \cite{wmd}, and also tune the following: For LSI and LDA the number of topics is searched in $ \{10,...,1000\}$. For BM25 $k_1 \in \{1.0,...,2.0\}$ and $b \in \{0.5,...,1.0\}$. For SIF $\alpha \in \{10^{-2},...,10^{-5}\}$. For CPTW and CPTW$_{\text{IDF}}$ $\tau \in \{0,...,1.0\}$. We use the best parameters found on the validation set when evaluating on the test set.

We use word embedding pretrained by the word2vec skip-gram model on Google News data\footnote{\url{https://code.google.com/archive/p/word2vec/}}.
For reproducibility purposes we use the gensim and scikit-learn python libraries\footnote{\url{https://radimrehurek.com/gensim/} and \url{https://scikit-learn.org/stable/}} for all compared methods, and use the implementation of SIF provided by the authors\footnote{\url{https://github.com/PrincetonML/SIF}}. 
We preprocess the datasets to remove non-alphanumeric characters, to tokenize text into lower cased word-tokens, and to remove stop words (based on the list by Salton \cite{salton1971smart}) as per \cite{wmd}. 

\begin{table*}[h]
    \centering
\scalebox{0.85}{
\begin{tabular}{|c|c|c|c|c|c|c|c|c|c|c|}
\hline
F$_1$ & BM25 & BOW & TF-IDF & LDA & LSI & WE-AVG & WMD & SIF & CPTW & CPTW$_{\text{IDF}}$\\ 
  & $\Scale[0.7]{\text{micro}\;\;\text{macro}}$  &
  $\Scale[0.7]{\text{micro}\;\;\text{macro}}$  &
  $\Scale[0.7]{\text{micro}\;\;\text{macro}}$  &
  $\Scale[0.7]{\text{micro}\;\;\text{macro}}$  &
  $\Scale[0.7]{\text{micro}\;\;\text{macro}}$  &
  $\Scale[0.7]{\text{micro}\;\;\text{macro}}$  &
  $\Scale[0.7]{\text{micro}\;\;\text{macro}}$  &
  $\Scale[0.7]{\text{micro}\;\;\text{macro}}$  &
  $\Scale[0.7]{\text{micro}\;\;\text{macro}}$  &
  $\Scale[0.7]{\text{micro}\;\;\text{macro}}$  
   \\ \hline
bbcsport & $.970\;\;.968$ & $.980\;\;.979$ & $\textbf{.986}\;\;\textbf{.987}$ & $.864\;\;.864$ & $.976\;\;.979$ & $.915\;\;.917$ & $.976\;\;.978$ & $.927\;\;.931$ & $.980\;\;.979$ & $.985\;\;.986$\\ \hline
   
twitter & $.711\;\;.470$ & $.715\;\;.522$ & $.728\;\;.540$ & $.642\;\;.353$ & $.721\;\;.508$ & $.721\;\;.537$ & $.704\;\;.445$ & $.714\;\;.495$ & $.724\;\;.532$ & $\textbf{.732}\;\;\textbf{.550}$\\ \hline

classic & $.849\;\;.860$ & $.935\;\;.935$ & $.947\;\;.950$ & $.840\;\;.845$ & $.921\;\;.920$ & $.926\;\;.933$ & $\textbf{.963}\;\;\textbf{.964}$ & $.931\;\;.937$ & $.955\;\;.958$ & $.957\;\;.960$\\ \hline

amazon & $.866\;\;.866$ & $.892\;\;.893$ & $.899\;\;.898$ & $.762\;\;.761$ & $.834\;\;.832$ & $.915\;\;.915$ & $.909\;\;.910$ & $.914\;\;.914$ & $.899\;\;.900$ & $\textbf{.918}\;\;\textbf{.918}$\\ \hline

reuters & $.925\;\;.859$ & $.894\;\;.818$ & $.873\;\;.814$ & $.916\;\;.833$ & $.955\;\;.864$ & $.956\;\;.885$ & $\textbf{.961}\;\;\textbf{.910}$ & $.958\;\;.888$ & $.950\;\;.877$ & $.942\;\;.894$\\ \hline


20news & $.467\;\;.470$ & $.650\;\;.649$ & $.661\;\;.662$ & $.488\;\;.482$ & $.565\;\;.572$ & $.641\;\;.634$ & $-\;\;-$ & $.644\;\;.636$ & $.661\;\;.659$ & $\textbf{.703}\;\;\textbf{.699}$\\ \hline

wiki & $.392\;\;.148$ & $.396\;\;.139$ & $.395\;\;.135$ & $.389\;\;.156$ & $\textbf{.417}\;\;.207$ & $.407\;\;.213$ & $-\;\;-$ & $.409\;\;.214$ & $.407\;\;\textbf{.217}$ & $.403\;\;.173$\\ \hline

\hline all$\ominus$WMD  & .740$\;\;$.663 & .780$\;\;$.705	& .784$\;\;$.712	& .700$\;\;$.613	& .770$\;\;$.697	& .783$\;\;$.719	& $-\;\;-$	& .785$\;\;$.716	& .797$\;\;$.732	& \textbf{.806}$\;\;$\textbf{.740} \\ \hline
all$\oplus$WMD  & .864$\;\;$.805	& .883$\;\;$.829	& .887$\;\;$.838	& .805$\;\;$.731& .881 $\;\;$.821	& .887$\;\;$.837	& .903$\;\;$.841	& .889$\;\;$.833	& .902$\;\;$.849	& \textbf{.907}$\;\;$\textbf{.862} \\ \hline

\end{tabular}
}
\caption{Micro and macro F$_{\text{1}}$. all$\ominus$WMD is the average score for all datasets, and all$\oplus$WMD is the average score for all datasets where WMD could be run. Bold denotes best scores.\label{f1roc} }
\vspace{-20pt}
\end{table*}

\subsection{\label{performancecomp}Findings}
Table \ref{f1roc} displays the performance (micro and macro $\text{F}_1$) of all methods on the 7 datasets. WMD scores are not shown for the \emph{20news} and \emph{wiki} datasets, because, after running WMD for 5 days, we stopped experiments with that model on the grounds of its poor efficiency. 
Due to this, in Table \ref{f1roc} we report the average scores separately for all datasets (all$\ominus$WMD) and for the 5 datasets that WMD was run on (all$\oplus$WMD).

CPTW$_{\text{IDF}}$ is the best performing method on both micro and macro $\text{F}_1$ on average across all datasets. CPTW is the second best performing method as per micro $\text{F}_1$ on all datasets, while WMD is the second best on the 5 datasets it could be run on. With respect to macro $\text{F}_1$, CPTW is the second best performing method on all datasets, and with WMD being the second best when only considering the 5 datasets WMD could be run on. When considering all datasets on macro $\text{F}_1$, then SIF was the best of the baselines, but with both CPTW and CPTW$_{\textrm{IDF}}$ performing better.
The worst performing method is LDA, which is most likely because LDA works better on texts with a large number of words. Indeed LDA performs relatively well on \emph{wiki} -- the dataset with the longest texts on average -- and noticeably worse than the other methods on most of the other datasets.



\subsection{Influence of $\tau$\label{sec:influence-tau}}
We further investigate the stability of the threshold $\tau$ of our model. Specifically, we investigate to what extent (or if at all) optimal values of $\tau$ correlate with a large inter versus intra class ratio in the embedded neighbourhood of a sample document (the intra versus inter class ratio indicates classification difficulty as explained next). We conduct this analysis only with CPTW$_{\text{IDF}}$, our best performing method in Table \ref{f1roc}.

We define \textit{inter vs intra class ratio} (IICR) as the ratio of average Euclidean distances from each point to its closest inter and intra class neighbours. Intra class neighbours refer to the points close to each other within the same class, and inter class neighbours refer to the points close to each other across different classes. The ratio of average Euclidean distances for the inter and intra class neighbours shows how similar the points from one class are compared to the other classes.
We study how IICR varies when $\tau$ is changed, and compute IICR as an average over all classes in a dataset for a specific $\tau$ as follows:
\begin{align}
\text{distance}_{c_1}^{\textrm{inter}} &= \frac{1}{|c_1|} \sum_{v_1 \in c_1} \; \sum_{v_2 \in N_{k}(v_1, C-\{c_1\})} || v_1 - v_2||_2 \\
\text{distance}_{c_1}^{\textrm{intra}} &= \frac{1}{|c_1|} \sum_{v_1 \in c_1} \; \sum_{v_2 \in N_{k}(v_1, c_1)} || v_1 - v_2||_2 \\
IICR &= \frac{1}{|C|} 
\sum_{c_1 \in C} 
\frac{ \text{distance}_{c_1}^{\textrm{inter}}}
{ \text{distance}_{c_1}^{\textrm{intra}} } \label{pointfraction}
\end{align}
where $C$ represents the set of classes, and $N_{k}(v_1,c_1)$ is the k closest points to $v_1$ from the $c_1$ class. An IICR score close to 1 means that inter and intra class samples are highly similar and thus hard to correctly classify. A higher IICR score than 1 means that inter and intra class samples are more dissimilar, which makes classification fundamentally easier. We consider the closest points, instead of all points, because the space should be transformed such that points become more similar to close intra class points, and become more dissimilar to the closest inter class points. We expect a large value of IICR to correlate well with an appropriate value of $\tau$ for a specific dataset, because we reason that our model should disambiguate semantically related and unrelated documents better. We choose the best performing k on each dataset.

\begin{figure*}[h]
\hspace{-5pt}
\includegraphics[width=1\linewidth]{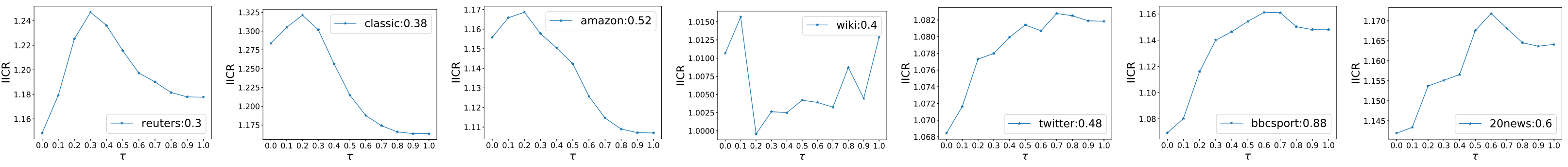}
\hspace{-5pt}
\vspace{-10pt}
\caption{\label{fig:robustness}Each graph shows the IICR as a function of $\tau$. The legend shows the best (average) $\tau$ from validation. 
}
\vspace{-10pt}
\end{figure*}


We compute the IICR for all datasets when varying $\tau$ and plot it with varying $\tau$ (Fig. \ref{fig:robustness}), where the legend in the graphs shows the optimal averaged $\tau$ across the cross validations per dataset. 
Generally, the graphs show that maximum ratios are correlated with $\tau$ values close to the optimal found in the cross validation, except for \textit{wiki}. This follows our intuition of our model being able to disambiguate the documents such that intra class samples become more similar compared to inter class samples. This means that our approach was able to make documents of the same class more similar when compared to other classes.
For \textit{wiki} most $\tau$ values perform very similar for the best $k$, and when considering the ratios we see a similar trend where all ratios are nearly identical.

\section{Conclusion}
\label{s:conc}
We presented \emph{Contextually Propagated Term Weights}, or \emph{CPTW} that propagate 
the weight of a word (computed via embeddings) to words occurring in similar contexts. The redistribution of weight has the effect of generalizing the semantics of a text, leading to improved discriminative power.
CPTW has low computational cost: state of the art word embeddings are used that can be precomputed efficiently offline, and the propagation itself can be performed efficiently using sparse matrix multiplications based on a precomputed matrix of word similarities.

Experimental evaluation against 8 baselines on 7 well-known datasets shows that CPTW yields the best micro and macro F1 scores on average across all datasets. Most notably, CPTW outperforms strong embedding based methods such as word mover's distance (WMD) \cite{wmd} and smooth inverse filtering (SIF) \cite{sif}; likewise, the experiments show that CPTW's notion of embedded neighbourhood is robust across datasets of different inter vs intra class ratio (i.e. across datasets of increasing classification difficulty). Using the ratio of average euclidean distance between inter and intra class kNN samples as a measure of classification difficulty of a dataset, we found that the value of the (sole) parameter $\tau$ in CPTW chosen by cross validation was close to the $\tau$ maximizing this ratio.

Our work complements recent efforts to employ word embeddings in novel ways that are both (i) computationally efficient, and (ii) semantics-sensitive in that discriminative power is increased by exploiting semantic \cite{rekabsaz2016generalizing,rekabsaz2017exploration,WangLCSL19} and structural \cite{HansenHASL19} similarities. Along these lines, future research directions include investigating whether, and to what extent, multiple similarity thresholds may improve the overall discriminative power of the method.


\begin{acks}
Partly funded by Innovationsfonden DK, DABAI (5153-00004A).
\end{acks}

\bibliographystyle{ACM-Reference-Format}

\end{document}